\documentclass[pra,showpacs,floatfix,amsmath,amsfonts,
twocolumn
]{revtex4}
\usepackage{graphics}
\usepackage{epsfig}
\usepackage{xspace}
\usepackage{soul}
\usepackage{color}

\newcommand{\PT}{{\cal PT}}

\newcommand{\be}{\begin{equation}}
\newcommand{\ee}{\end{equation}}

\newcommand{\sech}{\mathrm{sech}}
\newcommand{\arctanh}{\mathrm{arctanh}}
\newcommand{\arccoth}{\mathrm{arccoth}}

\begin{document}

\title{Stationary
			 through-flows in a Bose-Einstein condensate with a $\PT$-symmetric impurity }

\author{Dmitry A.  Zezyulin$^{1}$,
I. V.  Barashenkov$^{2,3,4}$, and Vladimir V.  Konotop$^{1}$}

\affiliation
{$^{1}$ Centro de F\'isica Te\'{o}rica e Computacional and Departamento de
F\'isica, Universidade de Lisboa, Campo Grande 2, Edif\'icio C8,  Lisboa 1749-016, Portugal \\
$^2$
Department of Mathematics,
University of Cape Town, Rondebosch 7701, South Africa
\\
$^3$ Joint Institute for Nuclear Research, Dubna 141980, Russia \\
$^4$ Department of Physics, University of Bath, Claverton Down,
Bath BA2 7AY, UK}
	
\date{\today}

\begin{abstract}
Superfluid currents in the boson condensate with a
 source and sink of particles are modelled by
 the
 $\PT$-symmetric Gross-Pitaevskii equation with a complex potential.
 We demonstrate the existence of through-flows of the condensate ---  stationary states with the asymptotically nonvanishing flux.
The through-flows come in two broad varieties
determined by the form of their number density  distribution. One variety is described by
 dip-like solutions featuring  a localised density depression; the other one comprises
   hump-like structures with a density spike in their core.  We exemplify each class by exact closed-form solutions.  For a fixed set of parameters of the $\PT$-symmetric potential,
stationary through-flows form continuous families parametrized by the strength of the background flux.
All hump-like and some dip-like members of the family are found to be stable.
We show that
the through-flows can be controlled by varying the gain-and-loss amplitude  of the complex potential and that
 these amplitude variations may produce
an anomalous  response of the flux across the gain-loss interface.

\end{abstract}
\maketitle

\section{Introduction}

  An important macroscopic  property of degenerate quantum gases is their ability to conduct superfluid currents.
 The superfluid flows in atomic Bose-Einstein condensates (BECs) have been thoroughly analysed in  a variety of settings characterised by real and complex
 external potentials~\cite{BECs}.
  The conservative systems modelled by the real
 potentials
  comprised  linear~\cite{OL} and nonlinear~\cite{NonlinLatt}
  optical lattices, localised impurities~\cite{impurities},  and potential barriers~\cite{barriers}.
   The superfluid currents in localised dissipative potentials  were also studied in atomic BECs, theoretically~\cite{Zeno_theor} and  experimentally~\cite{Zeno_experiment},
   \textcolor{black}{with the macroscopic Zeno effect~\cite{SK} being observed.} The corresponding physical contexts included  also  the exciton-polariton condensates~\cite{polariton}.

   In this paper, we consider hydrodynamic currents in a general nonconservative model with a complex potential~\cite{Miga_Rev}.
   Specifically, we are interested in parity-time ($\PT$-) symmetric arrangements.
   The
importance  of the $\PT$ symmetry for different branches of physics has been widely appreciated \cite{review} since it was introduced in the non-Hermitian quantum mechanics~\cite{BenderBout}.
   The quantum $\PT$-symmetric systems  are defined by non-Hermitian Hamiltonians commuting with the  composition of the space and time inversion.
Non-Hermitian $\PT$-symmetric Hamiltonians may have all-real spectra in certain parameter domains.

The similarity between the Schr\"odinger equation of quantum mechanics and the mean-field Gross-Pitaevskii equation inspired studies of BECs in the presence of $\PT$-symmetric potentials~\cite{BECs_PT,review}.
	Nontrivial  flows in  a $\PT$-symmetric condensate may result from the balance of the particle influx and leakage.  The persistent currents in $\PT$-symmetric atomic condensates have been considered primarily in the tight-binding approximation \cite{Cartarius}.
   This approximation reduces the rich wave dynamics of the  Gross-Pitaevskii equation to the interaction  of just a few  modes. As
   for the currents in the full infinite-dimensional  $\PT$-symmetric system, these
    have been discussed only recently, in connection with the jamming anomaly \cite{jamming} and
    nonlinear flows supported by
    pseudo-spectral singularities in a ring-shaped waveguide~\cite{ring_shape}.
       (The jamming anomaly consists
	in the drop of the flux from the pumped to the leaking site despite their common gain and loss amplitude being raised.
	It is a relative of  the macroscopic Zeno effect~\cite{SK,Zeno_theor,Zeno_experiment} understood as the attenuation of the condensate depletion as the atom removal rate
	is increased at the removal sites.)

 The paper \cite{jamming} focussed on localised stationary states, with
      all fluxes  assumed to be vanishing at  infinity.
As that earlier analysis,
the present study deals with stationary currents  in  the
Gross-Pitaevskii equation with a localised complex potential in the form of a  $\PT$-symmetric dipole. In difference to Ref.~\cite{jamming},
 we consider  the situation where in addition to  the flux generated by the dipole,
there are nonvanishing background fluxes --- that is, currents originating at  one infinity and running toward the other one.
We show that  the balance of the nonlinearity, dispersion and Hermitian well-shaped potential,
with the assistance of the  local flux generated by the gain-loss dipole,
produces stable  localised structures in  the condensate of constant density flowing at a constant speed.
The presence of gain and loss may give rise to highly nontrivial spatial profiles of the superfluid current.
In particular, we report stable stationary  structures consisting of a finite-size patch of current embedded in the background flow of opposite direction.

The outline of the paper is as follows.
 In the next section, the problem of stationary flows in the condensate with gain and loss is formulated
as a stationary $\PT$-symmetric Gross-Pitaevskii equation with
nonvanishing boundary conditions. In Sec.~\ref{sec:exact}
we establish the existence of the through-flow currents in a special type of
$\PT$-symmetric potential, the so-called Wadati potential.
The  structure of the Wadati potential allows to obtain
the  solution of the associated Gross-Pitaevskii equation in closed form.

Section \ref{sec:gauss} looks into stationary through-flows
  in generic $\PT$-symmetric potentials outside the Wadati variety.
  A potential with a  fixed set of parameters is shown to support
   a continuous family of through-flow solutions
    that can be parametrised by the asymptotic flux and background density.
    Part of the family have a density  dip  and the other part  density spike in their core.
Section \ref{stability}   examines stability of the through-flows.
    In section \ref{vary_gain} we explore the response of the interfacial flux associated with the stationary through-flows,
to the variation of the gain-loss coefficient.
Finally,  section ~\ref{sec:concl} summarises conclusions of this study.

	\section{Stationary through-flows}
	\label{sec:statement}
	
	
	We consider  the dimensionless Gross-Pitaevskii equation
	\begin{eqnarray}
		\label{evolut}
		i\Psi_t=-\Psi_{xx}+U(x) \Psi+2|\Psi|^2\Psi,
	\end{eqnarray}
	with a decaying potential  $U(x) \to 0$ as $|x| \to \infty$. The potential satisfies
	$U(x)=U^*(-x)$  and has the nature of a  $\PT$-symmetric defect.
	The sign of $\mathrm{Im} \, U(x)$ determines whether the particles are fed into the system (this is the situation in the region with $\mathrm{Im} \, U>0$)
	or eliminated from the condensate (the corresponding domain is characterised by $\mathrm{Im} \, U<0$).
		The nonlinearity in (\ref{evolut}) corresponds to the two-body repulsion between bosons.
	
	The quantity
	\[
	n(x,t)= |\Psi|^2
	\]
	gives the density of the number of particles in the condensate,
	\[
	\phantom{\frac12}   J(x,t)= i (\Psi_x^* \Psi- \Psi^* \Psi_x)
	\]
	is the associated flux, and
	\[
	v(x,t) = \frac{J}{n}
	\]
	is the superfluid velocity.

	The aim of this paper is to study nonlinear structures induced by a localized $\PT$ symmetric potential $U(x)$ in
	nonzero superfluid flows. Accordingly, 	we impose the boundary conditions  in the form
	\begin{subequations}
		\label{boundary}
		\begin{eqnarray}
			\lim_{x\to\pm\infty} n(x,t)&  = &  n_{\infty },    \label{BC1}  \\
			\lim_{x\to\pm\infty} J (x,t)&=&J _{\infty },
		\end{eqnarray}
	\end{subequations}
	where the background density $n_{\infty } \ne 0$ and flux
	$J_{\infty } \ne 0$.
		Solutions representing  stationary flows have the form
	$\Psi(x,t)=e^{-i\mu t} \psi(x)$, where $\mu$ is the chemical potential.
	The spatial part of the solution solves the equation
	\be
	-\psi_{xx} + U(x)
	\psi + 2\psi|\psi|^2 = \mu \psi.
	\label{A0}
	\ee
	In the stationary regime the number density, flux and superfluid velocity are time-independent:
	$n=n(x)$, $J=J(x)$, and $v=v(x)$.


	The stationary configurations of condensate  with  equal nonvanishing inflow and outflow {of particles}, will be referred to as the
	{\it through-flows\/} in what follows.

The absolute value of the background flux is uniquely
  determined by the  chemical potential
  and background density.
Making use of the boundary conditions \eqref{boundary} as well as the
 fact that $U(x) \to 0$, equation \eqref{A0} gives
\begin{equation}
\label{mu:Gauss}
\left(  \frac{J_\infty}{2n_\infty} \right)^2
=\mu-  2n_\infty.
\end{equation}

We also note
that for the given background density $n_\infty$
(and varying chemical potential),
 the background flux cannot exceed a certain finite limit.
Linearising equation \eqref{A0} with $U=0$ about the asymptotic solution
$ \psi= \sqrt{n_\infty} e^{ i (\nu /2) x }$, where
\be
 \nu =\frac{J_\infty}{n_\infty},     \label{vJn}
\ee
the small perturbation is found to have the form
\[
\delta \psi(x) = e^{i (\nu/2)  x}\left[ (A+iB)e^{-2\kappa x}+ (A^*+iB^*)e^{-2\kappa^* x} \right],
\]
where $A$ and $B$ are complex constants. The decay rate $2\kappa$ satisfies
\[
\kappa^2 \left[ \kappa^2- \left(  n_\infty- \frac{\nu^2}{4}   \right) \right]=0.
\]
Perturbations $\delta \psi$ that decay as $x \to \infty$ have
\[
\kappa^2=n_\infty-\frac{\nu^2}{4} >0.
\]
Substituting \eqref{vJn} in $\nu^2< 4n_\infty$ gives
\be
\left| J_\infty  \right|  \leq 2 \sqrt{n_\infty^3}. \label{bound}
\ee
We will be mindful of
this simple bound when considering the variation of  the background current $J_\infty$ (section \ref{sec:gauss}).

Our final remark here is that  solutions with nonvanishing condensate density
at infinity may only arise if $\mu>0$.  This is an obvious consequence of equation
\eqref{mu:Gauss}.

\section{Wadati potentials and exact solutions}
\label{sec:exact}

A simple way
 to demonstrate   the existence of   stationary  $\PT$-symmetric  through-flows
 and illustrate their properties,
  is to produce an explicit solution.
 To this end, we consider a $\PT$-symmetric potential of the special form \begin{equation}
\label{Riccati}
U(x)=-w^2(x) {+}iw_x(x),
\end{equation}
where $w(x)$ is an even real-valued function
(referred to as the {\it base\/} function in what follows).

  Wadati~\cite{Wadati}
was the one who has originally noted rich $\PT$-symmetric properties of potentials of this form.
The existence of localized nonlinear modes supported by potentials \eqref{Riccati} was established in~\cite{Abdul_Tsoi,ZK_Wadadti}. Ref.~\cite{J_Yang} studied  the symmetry breaking while Ref.~\cite{MMR} explored the  modulation instability of a constant-density solutions arising in the Wadati potentials.
  More complicated closed-form solutions in Wadati potentials were found in \cite{BZK}.
  In the present work, we exploit the analytical tractability of these potentials to derive an exact solution describing a stationary
  through-flow with a spatially localized  rarefaction or compression.

  \subsection{Two-parameter Wadati family}
  \label{2pW}

  To facilitate the analysis of the flux profile associated with a through-flow, we try to keep the corresponding density distribution simple.
  Specifically, we take
\begin{equation}
  \label{n(x)}
  n = \frac{\mu}{2} \left(1+a\, \sech^2y\right), \quad y=x\sqrt{\mu/2}\sin\phi.
  \end{equation}
  The density \eqref{n(x)}  is characterised by  two  parameters
    that can be freely chosen in the following ranges:
  \begin{equation}
   a> -1,  \quad  0 \leq  \phi \leq \pi/2.   \label{parameters}
 \end{equation}
 Negative values of $a$ ($-1< a<0$) correspond to densities with a  dip at the origin  and positive $a$  to hump-shaped distributions.
  The angle $\phi$  determines the characteristic width of the dip or hump.

The  Wadati potential (\ref{Riccati}) supporting the solution with the density \eqref{n(x)},  is generated by the  base function
   \be
   \label{w}
   w = -\frac{\sqrt{2\mu}}{4}\frac{2 \cos^2\phi  \cosh^2 y   + 3(a  +\sin^2\phi)} %
   { \cosh  y \sqrt {\cosh^2 y\cos^2\phi + a  +\sin^2\phi} }.
   \ee
   (See the Appendix for the derivation.)
   The
superfluid velocity $v=2\theta_x$ associated with the density \eqref{n(x)} and
potential \eqref{Riccati}+\eqref{w},
has the form
       \be
     \label{v}
  v  = -\frac{\sqrt{2\mu}}{2}\, \frac{\mathcal N}{\mathcal D},
     \ee
     with
   \[
     \mathcal N=   2  \cos^2\phi  \cosh^4 y  + (a + \sin^2\phi)(a + 3\cosh^2y)
   \]
   and
   \[
     \mathcal D=
      \cosh y   \left( a +  \cosh^2 y \right)\sqrt {\cosh^2 y\cos^2\phi+a+\sin^2\phi}.
   \]

The imaginary part of the Wadati potential with the base \eqref{w}
  is given by
 \be
 w_x= (a+\sin^2 \phi) \tanh y \,   {\mathcal W}(y).  \label{wx}
 \ee
 Here
 \[
 {\mathcal W}(y)=
 \frac{  \mu \sin \phi \left[ 4 \cos^2 \phi \cosh^2 y + 3(a+ \sin^2 \phi)\right]}
 {4 \cosh y   \, (\cos^2 \phi \cosh^2y + a + \sin^2 \phi)^{3/2}}
 \]
 is an even function; it is not difficult to verify that $ {\mathcal W}(y)>0$.
 Hence the derivative  \eqref{wx} is positive in one $x$-semiaxis and negative in the other.
     This arrangement corresponds to a $\PT$-symmetric dipole: the particles are gained in one half of the $x$-line and lost in the other one.

 Note that the ``polarity" of the dipole  switches around as the sign of the sum $a+ \sin^2 \phi$ is changed.
  When $a+ \sin^2 \phi<0$, the particles are gained in the left semi-axis and lost in the right one,
  whereas when $a+ \sin^2 \phi>0$, the region of gain is to the right of the region of loss.

\begin{figure}[t]
\begin{center}
	\includegraphics[width=1.0\columnwidth]{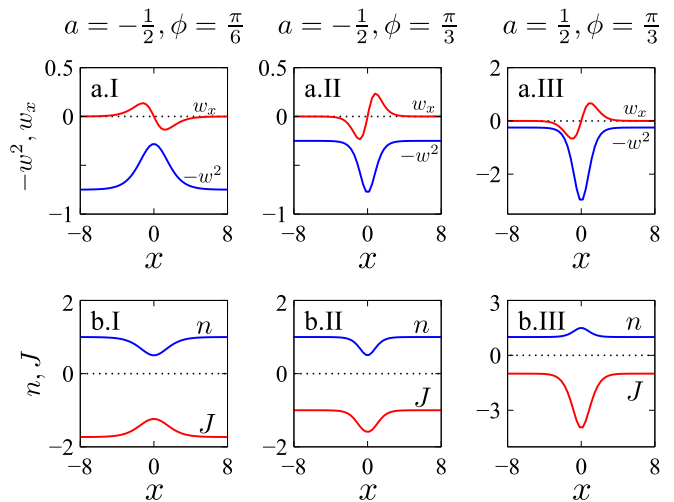}
	\caption{The Wadati  potentials \eqref{Riccati}, \eqref{w} (top row) and the corresponding exact solutions  \eqref{n(x)}, \eqref{v} (bottom row).
	The left column
	(column I) corresponds to $a=-1/2$, $\phi=\pi/6$;
	the middle column (II)  to  $a=-1/2$, $\phi=\pi/3$, and the right column
	(III)  to $a=1/2$, $\phi=\pi/3$.
	The three columns exemplify potentials and solutions at the points I, II, and III
	in Fig.~\ref{diagram}. In the top row, the blue  and red lines depict
	the real and imaginary parts of the potential \eqref{Riccati}, \eqref{w}.
	In the bottom row, the blue curves describe the number density and the red ones show the flux $J$
	associated with the solution \eqref{n(x)}, \eqref{v}.
	 In these plots, $\mu=2$.
	  }
	\label{examples}
	\end{center}
\end{figure}
\begin{figure}[!h]
	\begin{center}
		\includegraphics[width=.9\columnwidth]{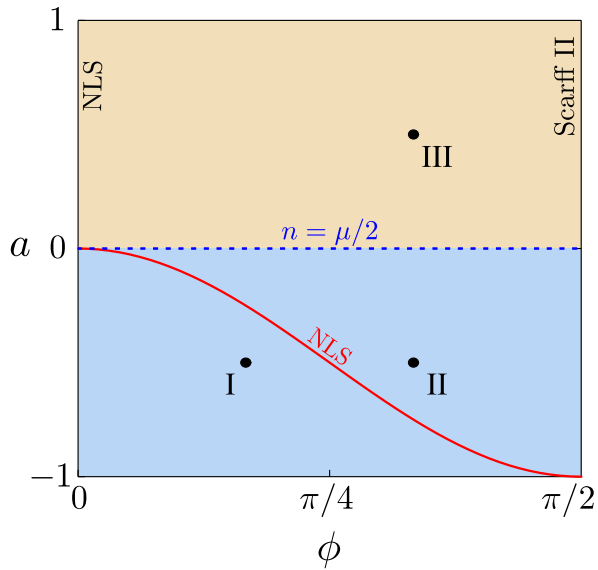}%
		\caption{(Color online) The parameter plane for the  Wadati potential \eqref{Riccati},  \eqref{w}
			and the corresponding
			solution \eqref{n(x)}, \eqref{v}.	In the regions I and II (coloured blue)
			the solution has the form of a dip in a constant density ---  see panels (b.I) and (b.II)
			of Fig.\ref{examples}.
			In the domain demarcated by III
			(tinted brown), the soliton is hump-like ---
			see panel (b.III)
			of Fig.\ref{examples}.	Note that  the domain III does not have an upper bound
			and extends indefinitely in $a$.
			The red curve separating domains I and II is given by $a=-\sin^2 \phi$.  Along this curve and
			along the vertical line $\phi=0$, the potential is real and the through-flow reduces to
			the nonlinear Schr\"odinger (NLS) dark soliton.
			Along the other boundary, $\phi=\pi/2$, the potential is the $\PT$-symmetric Scarff II.
		}
		\label{diagram}
	\end{center}
\end{figure}

The real and imaginary parts of the resulting potential $U$ are exemplified in Fig.~\ref{examples}, in the top row.
\textcolor{black}{The condensate density \eqref{n(x)} and flux $J=vn$ with $v$ as in
 \eqref{v},
 are shown in the bottom row of Fig.~\ref{examples}.
}

   Although the potential $U(x)$  does not decay to zero as $|x| \to \infty$,
 it has  finite and equal asymptotic values $U(\pm \infty)=U_\infty <0$.
 Accordingly,
  the ``uplifted" potential $\widetilde{U}(x)= U(x)-U_\infty$ does decay to zero and
  the analysis of the previous section remains valid if we replace $\mu$ with
   $\widetilde{\mu}=\mu-U_\infty$.



Computing the superfluid current for the solution (\ref{n(x)}), (\ref{v}), we  arrive at
\begin{equation}
J_\infty  =-\sqrt{ \frac{\mu^3}{2} } \cos \phi.   \label{ff}
\end{equation}
The background flux
 is negative for  all $\phi$ between $0$ and $\pi/2$.
This implies that  the condensate flows in the negative
  direction on the $x$-line --- irrespective of the choice of
  $\mu$, $a$, and $\phi$.
The \textcolor{black}{direction}  of the asymptotic current is unaffected even when the sign of $a+\sin^2 \phi$ is changed
so that  the local gain and loss sites are swopped around.
  This is clearly visible
in panels I and II of Fig.~\ref{examples}.
[The invariance of the current direction
 does not mean that there are no through-flows with positive asymptotic flux though.
 The positive through-flows result from expressions \eqref{alpha4} in the Appendix.
 Those have  $n(x)$ as in  \eqref{n(x)} but $w(x)$ and $v(x)$ opposite to the expressions in \eqref{w} and $\eqref{v}$.]

The density and  velocity fields being time-independent,
  the solution (\ref{n(x)}), (\ref{v}) represents a stationary mode pinned on a localised defect.
  In the frame of reference moving  with the  negative velocity $\nu= J_\infty/n_\infty$,
  the mode and defect  appear to travel at  a constant positive velocity $ -\nu$.
  The inequality  \eqref{bound} amounts to $-\nu<c$,
  where $c$ is  the speed of sound in the frame of the flow: $c=2\sqrt{n_\infty}$.
  This explains the absence of any  Cherenkov radiations excited by the moving defect.

Using  equations \eqref{n(x)} and \eqref{v},   we obtain the flux at the point $x=0$  separating the active from the lossy region:
\begin{equation}
J(0) =
 -\sqrt{\frac{\mu^3(a+1)}{8}}(a+\sin^2\phi+2).
\label{j0}
\end{equation}
The flux across the gain-loss interface  is negative  --- regardless of the choice of
  $\mu$, $a$, and $\phi$.  \textcolor{black}{
  When the potential parameters satisfy $a+\sin^2 \phi<0$,
  the flux negativity could seem to contradict  the fact that particles are gained to the left and lost to the right of the origin.
}

 \textcolor{black}{
  To resolve this ``paradox"  we  notice that
  	\be
  	J(0)  =-2  \int_0^\infty |\psi|^2  \,  \mathrm{Im}  \, U(x)   \, dx+J_\infty.
  	\label{js}
  	\ee
	The first term on the right accounts for the gain or loss of particles in the region $x >0$.
	Despite this term being positive for $a+\sin^2 \phi<0$, the flux $J(0)$ ends up being
	negative due to the second, negative, term.
  The fact that $J(0)<0$ implies simply that the
  background flux $J_\infty$ is greater, in absolute value,  than the gain or loss rate in the finite part of the system.
}

\subsection{Four special cases}
  \label{particular_cases}

  The  solution (\ref{n(x)}), (\ref{v})  and the underlying potential  (\ref{Riccati}), (\ref{w})
  include several particular cases that are relevant in the context of recent studies.
  The corresponding parameter pairs are marked in the parameter plane of Fig.\ref{diagram}.

 First, letting $a=0$ produces a solution of uniform density:
 \[
 \psi(x) = \sqrt{\frac{\mu}{2}}  \exp \left\{ i   \int  w(x) dx \right\}.
 \]
 The modulation instability of a solution of this sort
was studied in the equation with a periodic potential~\cite{MMR}.

Second, equations  (\ref{Riccati}) and (\ref{w})
 with  $\phi=\pi/2$
 amount to the familiar  $\PT$-symmetric Scarff~II potential~\cite{Scarff}.
 The corresponding base function is given by
\begin{equation}
w(x) = -\frac{3\sqrt{2}\mu}{4} {\sqrt{1+a}}\,\sech \, y.
\end{equation}
Solutions of the repulsive nonlinear Schr\"odinger equation with the
$\PT$-symmetric Scarff~II potential  and vanishing boundary conditions,
have been  studied in~\cite{Scarff_nonlin}. Our solution \eqref{n(x)}, \eqref{v}
with $\phi=\pi/2$ is different
from those localized modes. It represents a dip (for $-1<a<0$) or a hump (for $a>0$) over
a nonzero background (with an asymptotically vanishing flux).

Third, the choice $a = -\sin^2\phi$ corresponds to
 the spatially-homogeneous conservative
situation. In this particular case, the base
 function reduces to a constant,
\begin{equation}
w(x) \equiv w_0, \quad w_0 = - \sqrt{\frac{\mu}{2}}\cos\phi,
 \end{equation}
 so that the real part of the potential $U(x)$ becomes constant and the
 imaginary part  zero.
The solution (\ref{n(x)}), (\ref{v}) becomes the
   dark soliton of the repulsive nonlinear Schr\"odinger equation.

   The role of the curve $a=-\sin^2 \phi$ on the parameter plane of Fig.\ref{diagram}
   can be appreciated by evaluating the difference between the interfacial and background flux
   for the   solution (\ref{n(x)}), (\ref{v}). We have
   	\be
	|J(0)|-|J_\infty|= (a +\sin^2 \phi)
	\mathcal Q,  \label{dif} \ee
	where
	\[
\mathcal Q=	 \frac{\mu^{3/2}}{2 \sqrt2}
\frac{\sqrt{a+1} \cos \phi   +  a+3  }{\sqrt{a+1} + \cos \phi} >0. \]	In the region I (region below the curve), the interfacial flux is smaller than the background flux
in absolute value, while in the region II (above the curve), the difference \eqref{dif} is positive.
	[Compare panels (b.I) and (b.II) of Fig.\ref{examples}.]

        This change of sign is a natural consequence of equation \eqref{js} which relates
        the difference \eqref{dif} to the loss rate in the right semiaxis.
        Crossing the curve $a=- \sin^2 \phi$ switches the gain and loss around ---
        see the expression \eqref{wx} for the imaginary part of the Wadati potential.
        At points lying on the curve, the system does not experience any gain or loss;
        hence
        the current is   spatially uniform. In particular, $J(0)=J_\infty$.

The fourth  simple particular case is defined by $\phi=0$.
As in the previous situation, the base function is a constant here:
\[
w = -\frac{\sqrt{2\mu}}{4}\frac{2+3a}{\sqrt{1+a}}.
\]
Hence the $\phi=0$ limit also corresponds to the homogeneous conservative nonlinear Schr\"odinger equation.
 In this case, equations (\ref{n(x)}), (\ref{v}) give an $x$-independent solution:
\begin{eqnarray}
n=\frac{\mu}{2}(1+a),\quad
v=-\frac{\sqrt{2\mu}(a+2)}{2\sqrt{a + 1}}.
\end{eqnarray}

\section{Continuous families}
\label{sec:gauss}

Like solitons in conservative systems, localised modes
(solutions decaying to zero at infinities) in nonlinear $\PT$-symmetric equations
form a continuous family for each set of parameter values of the model 
(see reviews \cite{review, bright}).
The aim of this section is to show that the stationary $\PT$-symmetric
through-flows  share this property.
Namely, a fixed $\PT$-symmetric potential supports a family of through-flows
parametrised by the background number density  {and}  flux.

\begin{figure}
	\includegraphics[width=1.0\columnwidth]{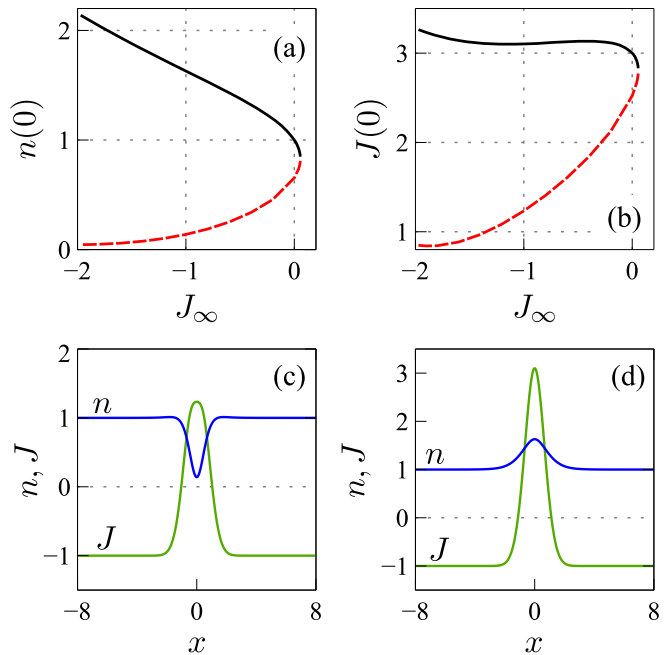}
	\caption{A
	family  of through-flows supported by the  $\PT$-symmetric
 Gaussian potential (\ref{eq:Gauss}) with $\gamma=3/2$ and $U_0 = 9/4$.
 The left panel  traces  $n(0)$ as $J_\infty$ is varied,  and  the right panel displays the corresponding  $J(0)$. Solid (black) and dashed (red) segments of curves correspond to stable and unstable solutions, respectively.
	 The bottom panels show
	 examples of a dip (c) and a hump (d) pertaining to $J_\infty=-1$.
	 The blue line depicts $n(x)$ and the green one shows $J(x)$.
	  In all panels, the background density is set to  $n_\infty=1$;
	  the largest value of $|J_\infty|$ attainable by this family is determined by
	  equation \eqref{bound}: $J_\infty=-2$. The starting point of continuation is the top point with
	  $J_\infty=0$;
	  here $n(0)$ and $J(0)$ are determined by the exact solution \eqref{eq:erf}: $n(0)=1$ and
	  $J(0)=3$.
		}
	\label{fig:families}
\end{figure}

We demonstrate the existence of continuous families of through-flows by
 considering a  {simple} $\PT$-symmetric potential in the form of a Gaussian potential well:
\begin{equation}
\label{eq:Gauss}
U(x) = -U_0\exp(-2x^2) -  2i\gamma x\exp(-x^2).
\end{equation}
Here $U_0>0$ is the depth of the well  and $\gamma$ is commonly referred to as the gain-loss coefficient, or the gain-loss amplitude.
A positive  $\gamma$ corresponds to the dissipative domain being placed to the right of the origin
and the active region being on its left. For negative $\gamma$ the arrangement is reversed:
particles are lost in the left and gained in the right  semiaxis.

\textcolor{black}{
Keeping  $U_0$, $\gamma$,
and the background density $n_\infty$ fixed,}
 we continue the solution $\psi(x)$ in $\mu$ numerically.
The background flux $J_\infty$ is related to $\mu$ by  equation \eqref{mu:Gauss}; it is instructive to
use $J_\infty$ as the parameter of the family.

Note that when $U_0=\gamma^2$,
 the gaussian potential (\ref{eq:Gauss}) belongs to the Wadati variety (\ref{Riccati}), with $w(x)=\gamma e^{-x^2}$.
  In this case,  equation \eqref{A0}
  admits  \cite{MMR}
  an exact uniform-density solution
\begin{equation}
\label{eq:erf}
\psi(x) = \exp \left\{ \frac{i  \sqrt{\pi} \gamma}{2} \,  \textrm{erf}\, (x) \right\}.
\end{equation}	
The background flux associated with
the solution \eqref{eq:erf}, vanishes:   $J_\infty=0$. Equation \eqref{mu:Gauss} gives the corresponding chemical potential: $\mu= 2 n_\infty$.
We use the above solution  as a starting  point for  our numerical continuation.

 A typical one-parameter family is traced in Fig.~\ref{fig:families}(a,b).
 In the left panel of that figure,
  each through-flow is represented by
 the density of the condensate at the origin.
 In the right panel,
we employ the interfacial flux as an alternative bifurcation measure.

Each bifurcation curve consists of two branches separated by a turning point.
Solutions on the whole of the red (dashed) branch and on a short section of the black (solid) branch characterised by positive $J_\infty$,
have the form of a dip in the uniform background: $n(0)< n_\infty$.
Solutions on the remaining part of  the black branch
($J_\infty < 0$)  represent humps: $n(0)> n_\infty$.
The largest attainable value  of $|J_\infty|$ in  Fig.~\ref{fig:families}(a,b)
is set by the  bound \eqref{bound}.

The dips and humps
are exemplified by Fig.~\ref{fig:families}(c,d).
The  density $n(x)$  and the superfluid flux $J(x)$ have the form similar to that of the dips and humps borne
by the Wadati potentials of the previous sections [cf. Fig.~\ref{examples}(b)].
The ``flat" solution \eqref{eq:erf} serves as a watershed between the two classes of through-flows.

\textcolor{black}{
For the most part of the bifurcation curve in  Fig.~\ref{fig:families}(b),   the
 background and  interfacial flux have opposite signs. \textcolor{black}{[This is at variance with the through-flow (\ref{n(x)}), (\ref{v}) supported by the Wadati potential (\ref{Riccati})+(\ref{w}).]} Therefore,
 each through-flow consists of a patch  of positive current  embedded in the
 negative background flow.
 Equation \eqref{js} and the inequality $J(0) J_\infty <0$
 imply  that the background flux $J_\infty$ is smaller in absolute value than the total loss rate
 in the  domain where atoms are removed, $x>0$. (The same  applies to the total gain rate  in the  domain where atoms are loaded, $x<0$.) In other  words,
 the  gain and loss corresponding to the chosen parameters of the potential
 are  intense enough to reverse the background flow locally.}
\textcolor{black}{Turning the background flux $J_\infty$ up} continuously
from negative to positive values does not \textcolor{black}{change} the direction of the interfacial current, so the above inequality is replaced with  $J(0) J_\infty >0$.
However this relation appears to be unsustainable over an extended
interval of $J_\infty$-values and the branch is forced to turn back
 soon after the background flux has changed its sign.

\section{Stable humps Unstable dips }
\label{stability}

To examine the stability of dips and humps, we let
\[
\Psi = \left[\psi + (f+ig) e^{\lambda t}
\right]e^{-i\mu t},
\]
where $\psi(x)$ is  the solution whose stability is analysed, and
$f(x)+ig(x)$ is  a small perturbation.
Substituting in  \eqref{evolut} and linearising in $f$ and $g$,
we arrive at an eigenvalue problem
\be
\mathcal H {\vec y} = \lambda \mathcal J {\vec y},
\label{EV}
\ee
where
\[
\mathcal H= \left(
\begin{array}{lr} \mathcal L
+ 6p^2+2q^2  &
  -\mathrm{Im} \,  U(x) + 4 pq \\
  \mathrm{Im} \,  U(x)  - 4 pq
  &
\mathcal L + 2p^2+6q^2
\end{array}
\right),
\]
\[
\mathcal L= - \frac{d^2}{dx^2} -\mu + \mathrm{Re} \, U(x),
\]
\[
{\vec y}= \left(
\begin{array}{c} f \\ g \end{array} \right),
\quad
\mathcal J= \left(
\begin{array}{lr}
0 & -1 \\
1 & 0
\end{array}
\right)
\]
and we have decomposed
\[
\psi(x)= p(x)+i q(x).
\]
The solution $\psi(x)$ is deemed unstable if there is at least one eigenvalue
with $\mathrm{Re} \, \lambda >0$.

The eigenvalue problem \eqref{EV} was analysed numerically.
The dip-like solutions on the red branch  in Fig.~\ref{fig:families} were found to be all unstable.
On the other hand,
 the black branch consists of  stable through-flows.
This  branch includes
all hump-like solutions and a small segment of dips (the segment with
 $J_\infty>0$).

The instability of the dips
 far from the turning point is due to a quadruplet of complex eigenvalues
 ($\pm \lambda,  \pm \lambda^*$)
 and a pair of opposite real eigenvalues.
  Approaching the turning point along the red curve, the complex eigenvalues converge, pairwise, on the imaginary axis --- but the real ones persist.
  At the turning point,
 the real pair converges at the origin and moves on to the imaginary axis.

\begin{figure}
		\includegraphics[width=\columnwidth]{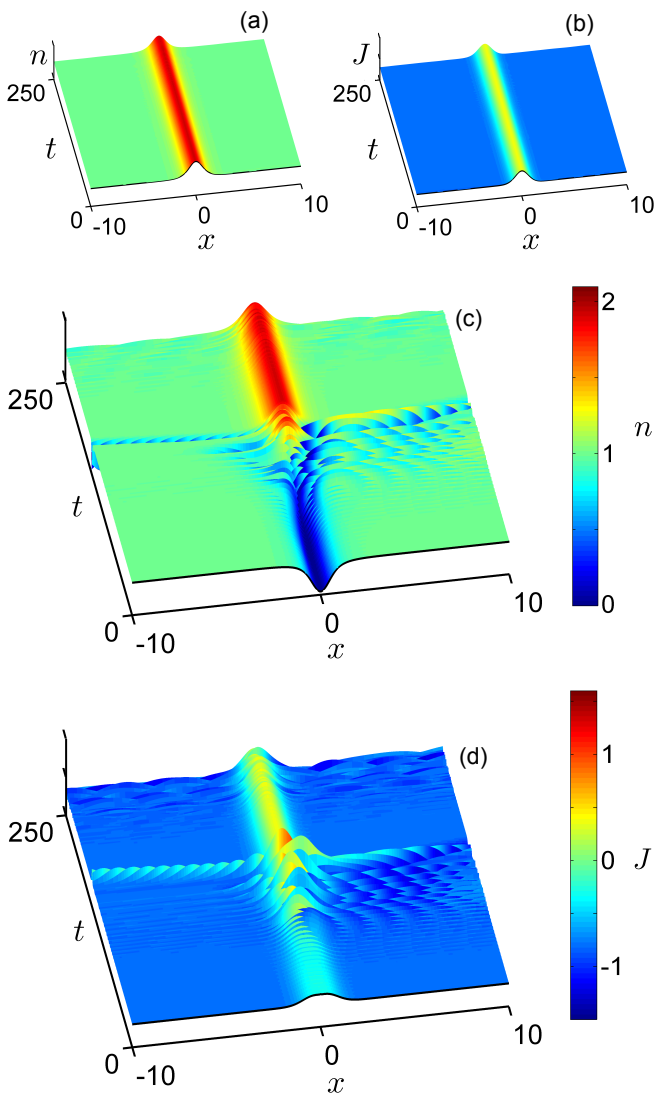}			
		\caption{The evolution  of a  hump (a,b) and  coexisting  dip  solution (c,d).
	The  panels (a,c) display the
	  density $n(x,t)=|\Psi|^2$, and the  panels (b,d) show the flux  $J(x,t) = 2(\arg\Psi)_x |\Psi|^2$.
	  The hump solution was initially perturbed by
	  a  gaussian-shaped perturbation with an amplitude of $2\%$ of the amplitude of the hump.
	  No explicit perturbation was added to the dip solution initially; here the instability was seeded by the discretisation  error.
	   In these plots,  $U_0=9/4$, $\gamma=0.4$ , $n_\infty=1$, and $J_\infty=-0.8$.	}
	\label{fig:evolution}
\end{figure}

The conclusions of the linear stability analysis
have been verified in direct computer simulations
 of the full time-dependent Gross-Pitaevskii  equation (\ref{evolut}).
   Fig.~\ref{fig:evolution} exemplifies the evolution of  perturbed  hump and dip solutions supported by the potential \eqref{eq:Gauss} with  the same values of $U_0$ and $\gamma$,
   and corresponding to the same  background flow.
 A small perturbation of the  hump is observed to  disperse away leaving the hump intact [Fig.~\ref{fig:evolution} (a,c)].
  In contrast, a small perturbation of the linearly unstable dip solution triggers its transformation into the hump with the equal background flux and density
  [Fig.~\ref{fig:evolution} (b,d)].
The emergent hump persists indefinitely.


\section{Varying  gain and loss}
\label{vary_gain}

In a structureless $\PT$-symmetric  dipole system, raising the gain and loss coefficient  boosts the flux between its active and dissipative components --- hence intensifying
the gain and loss in the dimer
\cite{jamming}.
In contrast, systems with internal spatial structure modelled by the Gross-Pitaevskii equation may exhibit an anomalous behaviour.
This  {\it jamming anomaly\/} manifests itself
  in the drop of the interfacial flux as a result of the growth of the gain-loss coefficient: $dJ(0)/ d\gamma<0$ \cite{jamming}.

In this section we explore the response of the interfacial flux associated with the stationary through-flows,
to the variation of the gain-loss coefficient.
For this analysis, we employ the complex potential  \eqref{eq:Gauss}.

\begin{figure}
		\includegraphics[width=\columnwidth]{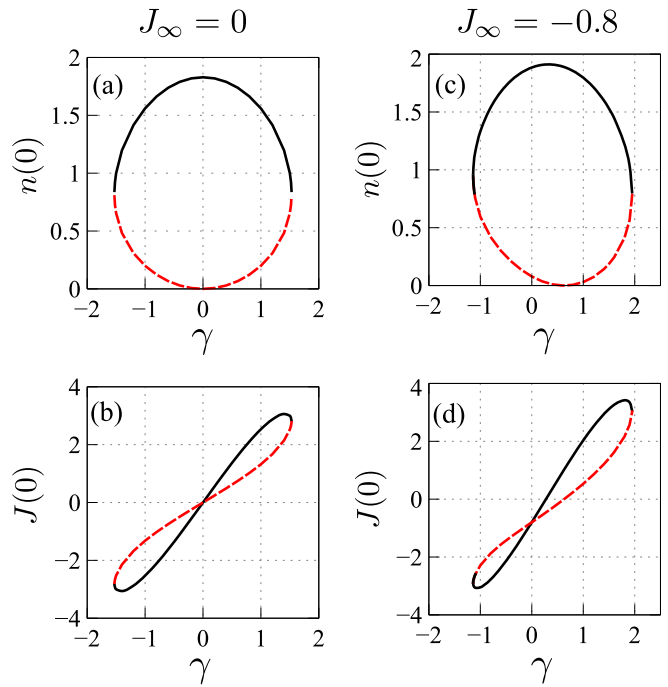}	
				\caption{The density at $x=0$ and the interfacial flux associated with the through-flows
	 in  the Gaussian $\PT$-symmetric potential (\ref{eq:Gauss}) with $U_0=9/4$ and varying $\gamma$.
	Panels (a) and (b) illustrate the case $J_\infty=0$. Panels (c) and (d) correspond to
	$J_\infty=-0.8$.
	Solid lines trace stable through-flows; unstable solutions are marked by dashes. In this figure, $n_\infty=1$.
		}
	\label{fig:branches}
\end{figure}

Choosing a pair of values for $n_\infty$ and   $J_\infty$, with
$|J_\infty| \leq 2 n_\infty^{3/2}$,
 fixes a particular member of the family of through-flows. Keeping the potential-well depth $U_0$ fixed and letting $\gamma$ to be varied, we
  determine this solution numerically. In Fig.~\ref{fig:branches} we represent this through-flow using
     two  scalar characteristics familiar from the previous section:
 $n(0)$ and $J(0)$.

The transformation $\gamma \to -\gamma$, $x \to -x$ leaves the equation
\eqref{A0} with potential  \eqref{eq:Gauss} invariant while flipping the sign of the flux $J$.
Accordingly, we restrict ourselves to nonpositive  values of $J_\infty$.
To recover a solution with $J_\infty>0$ and, say, $\gamma>0$,
one just needs to reflect a solution with $\gamma<0$ about the origin on the $x$-line.

Figure \ref{fig:branches} traces the family of  solutions with the vanishing background current  and
another one, corresponding to a  negative $J_\infty$.
In either case,  there are two coexisting through-flows  for each value
of $\gamma$.   As indicated by the solid and dashed lines in Fig.~\ref{fig:branches},
all hump-shaped solutions are stable whereas  most of the dip-shaped ones are unstable. The  loss of stability  occurs in the vicinity of the turning point.
(It would be tempting to believe that the stability is lost exactly at the turning point; however our numerics suggest
that this is not the case, at least for the nonzero asymptotic flux
$J_\infty$.)

 Away from the turning point, i.e.,  in the small-$|\gamma|$ region,
 the unstable dip solution features a quadruplet of complex
eigenvalues signalling an oscillatory instability. For larger $\gamma$, there also is
 a pair of opposite real eigenvalues.
As $\gamma$ approaches the turning point, the complex quadruplet converges on the imaginary axis
and then  the real pair moves on to the imaginary axis as well.

In the absence of the background flux  ($J_\infty=0$),  the   $n(\gamma)$-curve is symmetric with respect  to the $n$-axis (Fig.~\ref{fig:branches}(a))
   and the $J(\gamma)$-curve is symmetric under the $\gamma \to -\gamma, J \to -J$ reflection
   (Fig.~\ref{fig:branches}(b)).  This is an obvious consequence of the $\gamma \to -\gamma$, $x \to -x$ invariance of  equation
\eqref{A0}.
  The  interfacial flux $J(0)$ is positive if $\gamma>0$ and negative if $\gamma<0$. This  direction of the current  is set by the left-right arrangement of the gain and loss
  in \eqref{eq:Gauss}.

The  nonzero background flux  ($J_\infty<0$)
   breaks the symmetries of the diagrams --- see Fig.~\ref{fig:branches}(c,d).
  As a result, an interval of small positive $\gamma$ appears where     $J(0)<0$.
     For $\gamma$ in this interval, the background  flux $J_\infty$ is greater, in the absolute value,  than the gain or loss rate in the finite part of the system.
   [A similar situation arises in the Wadati potential \eqref{Riccati}, \eqref{w} with $a+ \sin^2 \phi<0$; see the discussion of the associated ``paradox"   at the end of section \ref{2pW}.]

   For $\gamma$ outside this interval adjacent to the origin, we have $J(0)>0$ ---  the gain and loss rates exceed the background flux.
   It is interesting to note
    two isolated values of $\gamma$, one for the hump and the other for the dip branch, where the interfacial flux vanishes: $J(0)=0$.
   At these $\gamma$, the background current is exactly compensated by the flux generated by the gain and loss.

 \textcolor{black}{  The dependence of the interfacial flux on the gain-loss amplitude is
     nonanomalous: $dJ(0)/d \gamma>0$,  for all $\gamma$ except small neighbourhoods of the turning points where   the flux jamming is observed:
$dJ(0)/d\gamma <0$.
   [See Fig.~\ref{fig:branches}(b,d).] }

 The nonmonotonic dependence of the interfacial current on the gain-loss coefficient in Figs.~\ref{fig:branches}(b, d)  is a demonstration of the {\em macroscopic Zeno effect}~\cite{SK,Zeno_theor,Zeno_experiment}. Indeed,
 the descending section of the curve on the $J(0)$ vs $\gamma$ diagram, with $J_\infty$ being fixed, corresponds to the suppression of the loss of the condensed atoms in the domain where they are removed (increase of number of atoms in the domain where they are loaded) as the intensity of removal (loading) of atoms is increased.

It is worth noting that
the domain of existence and stability  of the through-flows
can be controlled by varying    the background flux $J_\infty$.
For instance, the existence domain of solutions with $J_\infty=0$ is bounded from above by  the turning point at $\gamma_0 \approx 1.53$
[Fig.~\ref{fig:branches} (a,b)].
Raising the magnitude of the negative flux $J_\infty$ to $0.8$ shifts  the turning point   $\gamma_0$   to  approximately $1.95$  [Fig.~\ref{fig:branches} (c,d)].
Thus, sending  superfluid current through the system with gain and loss
ensures that the condensate  exists and remains stable  under  larger gain and loss amplitudes:
\[
\left. \phantom{\frac12}  \gamma_0 \right|_{J_\infty=-0.8} > \left.  \phantom{\frac12}  \gamma_0 \right|_{J_\infty=0}.
\]

Finally, we note that
the  hump- and dip-like solutions discussed in sections \ref{sec:gauss}-\ref{vary_gain}  are not the only through-flows supported by the $\PT$-symmetric system with background flux.
In particular, we were able to detect
branches of symmetric double- and triple-dip solutions in addition to the single-dip branch.
We do not elaborate on these here since all  multi-dip complexes were found to be unstable ---
like their  individual constituents.

\section{Concluding remarks}
\label{sec:concl}

In this paper,  we considered superfluid currents in the boson condensate with the  symmetrically arranged
source and sink of particles.  We have demonstrated the existence of  the  through-flows of the condensate ---  stationary states with the asymptotically nonvanishing flux.
These structures were described by solutions of the $\PT$-symmetric Gross-Pitaevski equation  with nonvanishing boundary conditions.

The stationary  through-flows were exemplified by explicit solutions
associated with the $\PT$-symmetric Wadati potentials. We have shown that the through-flows fall under two broad classes
determined by the form of their number density distribution.
The dip-like solution features a localised density depression and the hump-like structure has a density spike in its core.

The stationary through-flows supported by a given $\PT$-symmetric potential
 form a continuous family that can be parametrised by the asymptotic flux and background density.
The family includes both dip- and hump-like solutions.
All humps have been found to be stable whereas the  dip-like solutions are unstable --- except for a short interval of stability adjacent to the parameter domain of humps. \textcolor{black}{
It is important to emphasise that instability is an inherent property of the dip solutions that persists for any  arrangement  of gain and loss relative to the direction of the background flux.}

We have also explored the response of the  interfacial flux associated with the stationary through-flows to the variation of the gain-loss coefficient and described the associated anomalies.

\appendix

\section{Wadati potential and exact solutions}

In this Appendix, we derive the exact solution \eqref{n(x)}, \eqref{v}
of the stationary  Gross-Pitaevskii equation  with the Wadati base function \eqref{w}.

Following the procedure outlined in \cite{BZK}, we write  equation (\ref{A0}) as a first-order system
\begin{align}
\psi_x -  iw \psi + \phi=0, \nonumber \\
 \phi_x+   i w \phi +2 \psi|\psi|^2- \mu\psi=0.
 \label{E3}
\end{align}
The solution $\psi(x)$ and the base function $w(x)$ are  constructed in parallel.
The procedure is not assuming that $w(x) \to 0$ as $|x| \to \infty$
and accordingly, the Wadati potential \eqref{Riccati} does not have to approach zero at the infinities.

Substituting the polar decompositions
\[
\psi = \rho e^{i\theta},  \quad     \phi = \sigma e^{i\chi}
\]
in \eqref{E3} gives equations for the moduli and arguments of  $\psi(x)$ and $\phi(x)$:
\begin{align}
\rho_x + \sigma \cos \alpha=0, \label{R1} \\
\sigma_x - \rho(\mu- 2 \rho^2) \cos \alpha=0,  \label{R2} \\
(\theta_x-w)\rho + \sigma \sin \alpha=0,  \label{R3} \\
(\chi_x+w) \sigma + \rho(\mu-2 \rho^2) \sin \alpha=0. \label{R4}
\end{align}
Here
\[
\alpha = \chi -\theta
\quad (0 \leq \alpha \leq \pi).
\]
Eliminating $\cos \alpha$ between \eqref{R1} and \eqref{R2} we obtain
 a  conservation law~\cite{ZK_Wadadti}
\begin{equation}
\mu \rho^2 - \rho^4 + \sigma^2 = C.  \label{cl}
\end{equation}
Here $C$ is a  constant that can be determined by the boundary conditions.

The boundary condition \eqref{BC1} and the conservation law \eqref{cl} imply that $\rho(x)$ and $\sigma^2(x)$ tend to   constant values as $x \to \pm \infty$.
A simple (but not necessarily unique) option compatible with \eqref{R1} and \eqref{R2},
 is to assume that
 $\sigma (x) \to 0$ and $\rho^2(x) \to \mu/2$ as $|x| \to \infty$.
 The constant $C$ is then identified as
  $C=\mu^2/4$ and equation \eqref{cl} simplifies to
\begin{equation}
\label{eq:sigma}
\sigma = \pm (\mu/2 - \rho^2).
\end{equation}
 Here the top and bottom sign correspond to solutions with $\rho^2 \leq \mu/2$ and $\rho^2 \geq \mu/2$,
 respectively.

Making use of the relation (\ref{eq:sigma}), we express  $\theta_x$ and $w$  from  \eqref{R3} and \eqref{R4}:
\begin{eqnarray}
\label{eq:thetax}
\theta_x = -\frac{\alpha_x}{2} \mp \frac{\sin\alpha}{2\rho} \left(\frac{\mu}{2}+\rho^2\right),\\
\label{eq:w}
w = -\frac{\alpha_x}{2} \pm \frac{\sin\alpha}{2\rho}\left(\frac{\mu}{2} - 3\rho^2\right),
\end{eqnarray}
 while an equation for
 $\rho$  is straightforward from \eqref{R1}:
\begin{equation}
\label{eq:rhox}
\rho_x \pm (\mu/2-\rho^2)\cos\alpha=0.
\end{equation}

The general solution of (\ref{eq:rhox})  with the top sign ($\rho^2 \leq \mu/2$),  is
\begin{equation}
\label{eq:rhop}
\rho = -\sqrt{\frac{\mu}{2}} \tanh\left[\sqrt{\frac{\mu}{2}} \int \cos\alpha(x)d x\right],
\end{equation}
where the indefinite integral  incorporates the constant of integration.

Let
\be
\alpha (x) = \pi - \arccos \left(\frac{\sin\phi   \, \sinh y}{\sqrt{\cosh^2y+a}}\right),
\label{alpha1} \ee
where
\[
 y=x \sqrt{\mu/2}\sin\phi,
 \]
and $a$,  $\phi$ are two parameters:
$ -1 < a \leq 0$ and $0 \leq \phi \leq \pi/2$.
If we choose the constant of integration so that
\begin{equation}
\int \cos\alpha(x)dx = -\sqrt{\frac{2}{\mu}}\,\arctanh\sqrt{1+a \, \sech^2y}, \label{eq:intcos}
\end{equation}
we will ensure  that the  argument of $\tanh$ in \eqref{eq:rhop}  is negative,
hence the right-hand side of  \eqref{eq:rhop} is positive for all $x$. Thus,
equation  \eqref{eq:rhop} with the integral as in \eqref{eq:intcos}
defines the modulus of a solution to the stationary Gross-Pitaevskii equation.
The phase of this solution is  determined from
 (\ref{eq:thetax})  while
 (\ref{eq:w}) gives the  potential-base function
 of the equation \eqref{A0}. 

Turning to equation (\ref{eq:rhox})  with the bottom sign ($\rho^2 \geq \mu/2$),
its general solution  reads
\begin{equation}
\label{eq:rhom}
\rho = \sqrt{\frac{\mu}{2}} \coth\left[\sqrt{\frac{\mu}{2}} \int \cos\alpha(x)dx\right].
\end{equation}
This time, we let
\be
\alpha (x) =   - \arccos \left(\frac{\sin\phi \sinh y}{\sqrt{\cosh^2y+a}}\right),
\label{alpha2}
\ee
with
\[
y=x \sqrt{\mu/2}\sin\phi
\]
and two parameters,
$a \geq 0$ and  $0 \leq \phi \leq \pi/2$.
Choosing the integration constant so that
\begin{equation}
\int \cos\alpha(x)dx =  \sqrt{\frac{2}{\mu}}\,\arccoth\sqrt{1+a \, \sech^2y},   \label{R10}
\end{equation}
we make sure that
 the
 argument   of $\coth$ in   \eqref{eq:rhom} is             nowhere        negative.
 Consequently, the expression  \eqref{eq:rhom} with the integral as in \eqref{R10},
 gives the absolute value of another solution the Gross-Pitaevskii equation \eqref{A0}.
 Its phase $\theta(x)$ and the equation's base function
  $w(x)$  are straightforward from  (\ref{eq:thetax}) and
   (\ref{eq:w}).

The  absolute values \eqref{eq:rhop}+\eqref{eq:intcos}   and
\eqref{eq:rhom}+\eqref{R10},  can be written in a unified simple form --- see equation
\eqref{n(x)} in the main text. When $a<0$,
the number density $n=\rho^2(x)$ shows a dip at the origin, whereas
 positive values of $a$ pertain to a hump of the density.
 The corresponding superfluid velocities
$v=\theta_x$ and the potential base functions $w(x)$
also admit a simple expression; see \eqref{v} and  \eqref{w}, respectively.

Finally, we can choose $\alpha(x)$ as the negative of
expression \eqref{alpha1} or \eqref{alpha2}:
\begin{align}
\alpha= \arccos  \left( \frac{\sin \phi \,  \sinh y}{\sqrt{\cosh^2 y+a}} \right)  - \pi,  &
\quad -1 \leq a \leq 0, \nonumber  \\
\alpha= \arccos \left( \frac{   \sin \phi \,  \sinh y}{\sqrt{\cosh^2 y+a }} \right), &
\quad a \geq 0.
\label{alpha4}
\end{align}
The resulting solutions  will have the  sign of $\theta_x$
(and hence the sign of the asymptotic flux)  opposite to the one associated with \eqref{alpha1} and \eqref{alpha2}.
\\

\acknowledgments
IVB thanks Ricardo Carretero and Boris Malomed for instructive conversations.
The work of DAZ and VVK was supported by the FCT (Portugal) through
the grants UID/FIS/00618/2013 and PTDC/FIS-OPT/1918/2012.
IVB wishes to acknowledge financial support from
 the NRF of South Africa (grants UID 85751, 86991, and 87814) and  the European Union's Horizon
2020 research and innovation programme under the Marie Sk{\l}odowska-Curie grant agreement
No 691011.

\end{document}